\documentclass[twocolumn,twocolappendix]{aastex631}
\usepackage{amsmath}

\begin{document}

\title{Imaging the jet of MWC 349A with resolved Radio Recombination Line emission from ALMA}

\author[0000-0001-5191-2075]{Antonio Martínez-Henares}
\affil{Centro de Astrobiología (CSIC-INTA)\\
Ctra. de Torrejón a Ajalvir, km 4 \\
28850 Torrejón de Ardoz, Madrid, Spain}

\author[0000-0003-2384-6589]{Qizhou Zhang}
\affil{Center for Astrophysics $\vert$ Harvard \& Smithsonian \\
60 Garden Street \\
Cambridge, MA 02138, USA}

\author[0000-0003-4493-8714]{Izaskun Jiménez-Serra}
\affil{Centro de Astrobiología (CSIC-INTA)\\
Ctra. de Torrejón a Ajalvir, km 4 \\
28850 Torrejón de Ardoz, Madrid, Spain}

\author[0000-0003-4561-3508]{Jesús Martín-Pintado}
\affil{Centro de Astrobiología (CSIC-INTA)\\
Ctra. de Torrejón a Ajalvir, km 4 \\
28850 Torrejón de Ardoz, Madrid, Spain}

\author[0000-0002-2711-8143]{Nuria Huélamo}
\affil{Centro de Astrobiología (CSIC-INTA)\\
Ctra. de Torrejón a Ajalvir, km 4 \\
28850 Torrejón de Ardoz, Madrid, Spain}

\author[0000-0002-1082-5589]{Sirina Prasad}
\affil{Center for Astrophysics $\vert$ Harvard \& Smithsonian \\
60 Garden Street \\
Cambridge, MA 02138, USA}

\author[0000-0002-3882-4414]{James Moran}
\affil{Center for Astrophysics $\vert$ Harvard \& Smithsonian \\
60 Garden Street \\
Cambridge, MA 02138, USA}

\author{Alejandro Báez-Rubio}
\affil{UWC Mahindra College \\
Village Khubavali, PO Paud \\
MH 412 108, India}



\begin{abstract}
Jets and disk winds arise from materials with excess angular momentum ejected from the accretion disks in forming stars. How these structures are launched and how they impact the gas within the innermost regions of these objects remains poorly understood. MWC349A is a massive star that has a circumstellar disk which rotates in accord with Kepler's Law, with an ionized wind and a high-velocity jet launched from the disk surface. The strongly maser-amplified emission of hydrogen radio recombination lines (RRLs) observed toward this system provides a comprehensive picture of its ionized environment with exquisite detail. In this Letter, we present ALMA observations of the H26$\alpha$ RRL and continuum emission obtained with the highest angular resolution ever used toward this source (beam of $\sim$0.02"). The maser RRL emission is resolved for the first time and clearly delineates the ionized disk, wind and jet. We analyzed the RRL data cubes with the 3D non-LTE radiative transfer model MORELI, confirming that the jet is poorly collimated. We found that the jet orientation is closer to the rotation axis of the system than derived from spatially unresolved data. This study confirms that hydrogen RRL masers are powerful probes of the physical structure and kinematics of the innermost ionized material around massive stars. 
\end{abstract}



\section{Introduction} \label{sec:intro}

MWC$\,$349A is a well-known massive star \mbox{($\sim$20-30$\,$M$_{\odot}$)} located at $\sim$\mbox{1.2$\,$kpc} \citep{cohen1985} that has been the object of numerous studies across a wide wavelength range. This star is surrounded by an almost edge-on circumstellar disk \citep{danchi2001} in Keplerian rotation \citep{planesas1992,weintroub2008}. The radio continuum emission shows a 'X-shape' and flux density scaling of $\nu^{0.67}$ at centimeter and millimeter wavelengths \citep{tafoya2004}, well modeled as an ionized wind of an electron density that decreases with the radial distance as $r^{-2.14}$ and an isotropic mass loss expanding at a constant velocity \citep{olnon1975}. One of the distinctive aspects of MWC$\,$349A is the maser amplification of its hydrogen Radio Recombination Line (RRL) emission \citep{martinpintado1989}, which has been extensively observed across the radio wavelength spectrum since its discovery (see \citealt{baezrubio2013} for a review). The case of MWC$\,$349A has motivated the search and discovery of RRL masers in other massive systems such as Cep$\,$A$\,$HW2 \citep{jimenezserra2011}, Mon$\,$R2-IRS2 \citep{jimenezserra2013,jimenezserra2020}, $\eta$ Carina \citep{abraham2014,abraham2020}, MWC$\,$922 \citep{sanchezcontreras2019} or G45.47+0.05 \citep{zhang2019}.

Radio interferometric observations towards MWC$\,$349A with the Plateau de Bure Interferometer (PdBI, \citealt{martinpintado2011,baezrubio2013}), and the Submillimeter Array (SMA, \citealt{zhang2017}) revealed that the ionized wind is expanding radially and rotates in the same sense as the disk, removing angular momentum from the system. Recent high-sensitivity Atacama Large Millimeter/submillimeter Array (ALMA) observations have explored the highest velocities of the RRL emission for the first time, revealing a high-velocity jet in addition to the ionized disk and wind \citep{prasad2023,martinezhenares2023}. In all these works, the RRL emission is not spatially resolved and the relative spectroastrometry is conducted by determining the position of its 2D Gaussian centroid in each frequency channel. This method enables the determination of the mean location of the emission as a function of radial velocity with a positional accuracy that is inversely proportional to the SNR value \citep{condon1997}. The achieved angular precision is much higher than the synthesized beam size, typically of (sub-)au scales, due to the maser amplification. By comparing the radio continuum, RRL profiles and centroids with the non-LTE 3D radiative transfer MOdel for REcombination LInes MORELI \citep{baezrubio2013}, the physical structure and kinematics of the Keplerian disk and the ionized wind \citep{baezrubio2013,baezrubio2014} and of the jet of MWC$\,$349A \citep{martinezhenares2023} have been constrained. However, these components have neither been resolved nor been imaged directly before, leading to uncertainties in the determination of the structure and kinematics of the ionized gas in the model.

In this Letter, we report new ALMA observations that resolve for the first time the structure of the sub-mm continuum and H26$\alpha$ RRL emission toward MWC$\,$349A using ALMA's longest baseline configuration. We compare these observations with predictions from the MORELI code and find that a small modification in the orientation of the jet and the disk with respect to the model from \cite{martinezhenares2023} reproduces with high fidelity the resolved continuum and RRL emission. This proves the power of the MORELI code in providing an accurate picture of the physical structure and kinematics of the innermost ionized material around massive stars even when the emission remains unresolved.

\section{Observations and data reduction} \label{sec:obs}

The observations were performed with ALMA (Project code 2019.1.01069.S: PI Qizhou Zhang) on 2021 September 5 and 10 in the array configuration C43-9/10 (longest baseline of \mbox{16200$\,$m}). The spectral setup consisted of two spectral windows for the continuum centered at 340.610 and \mbox{342.506$\,$GHz} of \mbox{1.875$\,$GHz} bandwidth, and two spectral windows centered on the H26$\alpha$ RRL (\mbox{353.623$\,$GHz}) to obtain the line emission, one with higher spectral resolution than the other. The line spectral window of lower spectral resolution had a bandwidth of \mbox{1.875$\,$GHz} and channel width of \mbox{0.98$\,$MHz} or \mbox{$\sim$0.82$\,$km~s$^{-1}$}, while the other one was \mbox{0.469$\,$GHz} wide with a spectral resolution of \mbox{0.24$\,$MHz} or \mbox{$\sim$0.21$\,$km~s$^{-1}$}. The phase center for the MWC$\,$349A observations was $\alpha$(J2000)$=20^h32^m45^s.528$, $\delta$(J2000)$=40^{\circ}39'36".623$. Calibrators J2253+1608 and J2015+3710 were observed for bandpass and gain calibration, respectively. The initial calibration of the data was obtained from the ALMA pipeline. We performed self-calibration with CASA on the calibrated continuum and line data with the averaged continuum and the brightest channel of the line, respectively. The continuum self-calibration tables were obtained after five iterations of the phase calibration and one amplitude calibration, with a solution time interval of 16 seconds. We used the same solution interval to obtain the line self-calibration tables after five rounds of phase and one round of amplitude calibration. The continuum subtraction of the line spectral windows was done with the line-free channels of the wide spectral window. Finally, we imaged the data with CASA's routine \textit{tclean} using Briggs weighting with a \textit{robust} parameter of 0.0. The resulting synthesized beam size was of 0.025 $\times$ 0.012$\,$arcsec (30$\,$au $\times$ 14$\,$au, assuming a distance of 1.2$\,$kpc) with a position angle (PA) of $-$7$^{\circ}$.1 for the continuum, and same beam size with a PA of $-$3$^{\circ}$.3 and $-$11$^{\circ}$.7 for the narrow and wide spectral window line cubes, respectively. 

\section{Results and discussion} \label{sec:resultsdiscussion}

    \subsection{Continuum emission at 341 GHz} \label{sec:continuum}

    \begin{figure*}
            \includegraphics[width=\textwidth]{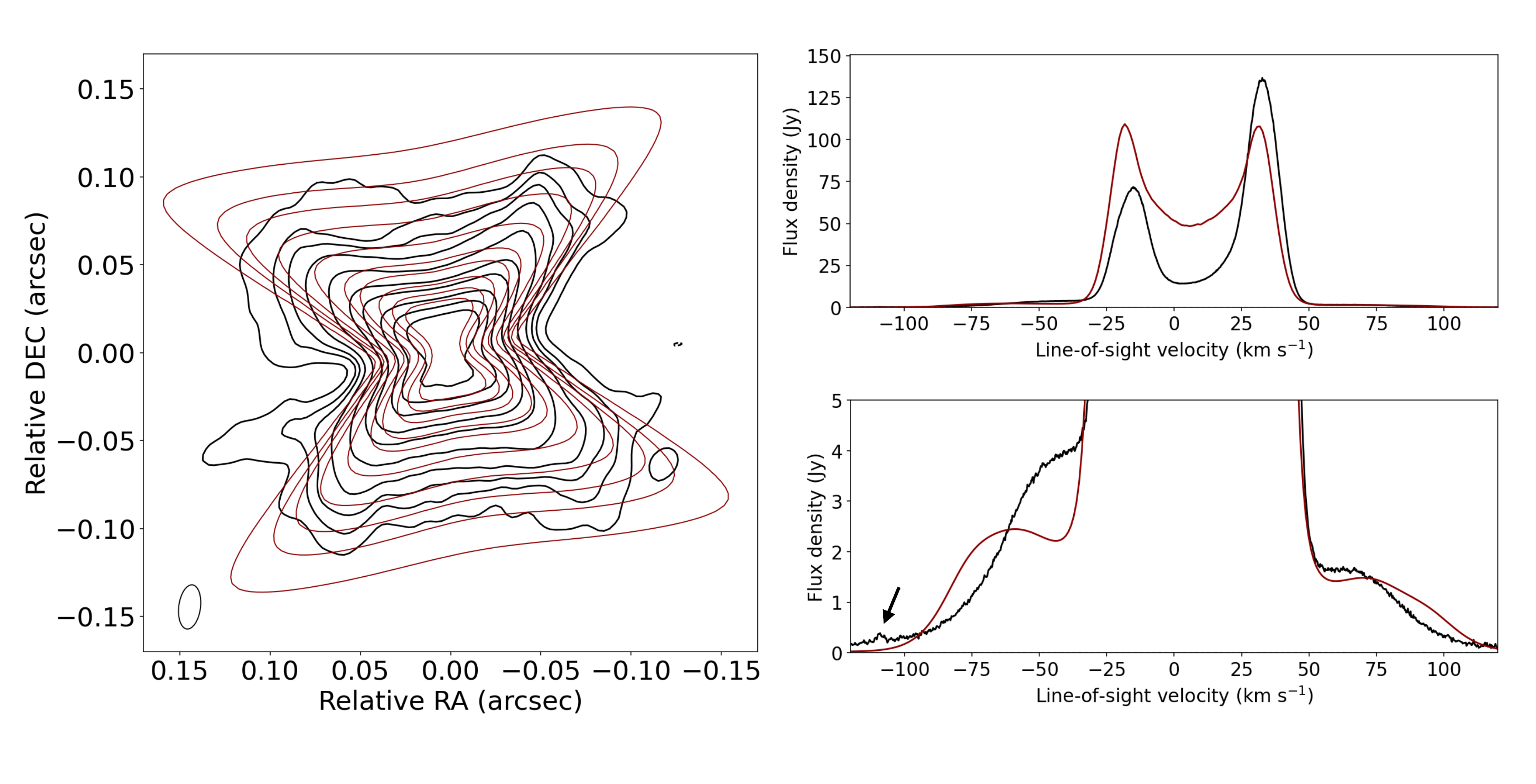}
            \centering
            \caption{\textit{Left panel:} Continuum emission imaged with ALMA at 341$\,$GHz toward MWC$\,$349A (black contours). The continuum emission modeled at 341$\,$GHz with the MORELI code is shown in red contours. Contour levels are \mbox{[$-$3,10,20,30,50,100,200,300,500,700,900]} times the 1$\sigma$ \textit{rms} noise level of the observations, 0.14$\,$mJy~beam$^{-1}$. The synthesized beam of size 0.025 $\times$ 0.012$\,$arcsec is represented with the ellipse at the bottom left corner. \textit{Top right panel:} Observed (black) and modeled (red) integrated spectral profile of the H26$\alpha$ line emission toward MWC$\,$349A. \textit{Bottom right panel:} Zoom-in into the lower intensity levels of the H26$\alpha$ spectral profile. The arrow indicates additional line emission, which is discussed in Sect. \ref{sec:newline}. \label{fig:continuum}}
    \end{figure*}

        The distribution of continuum emission is shown in black contours in the left panel of Figure \ref{fig:continuum}. This is the highest frequency in which the continuum emission toward MWC$\,$349A has been resolved. The total integrated flux measured is 2.4$\pm$0.2$\,$Jy, in agreement with the SMA and ALMA fluxes reported in \cite{prasad2023}. We assume the nominal 10\% uncertainty in the flux density measurement. The hourglass or 'X' shaped morphology seen at 2, 1.3, and 0.7$\,$cm with the VLA \citep{whitebecker1985,martinpintado1993,tafoya2004} and at 217$\,$GHz with ALMA \citep{prasad2023} is also observed in our ALMA continuum image at 341$\,$GHz. This suggests that the continuum emission is dominated by free-free radiation from a constant velocity expanding outflow \citep{olnon1975,panagiafelli1975} and not by thermal dust emission, which is also consistent with the spectral index of 0.64$\pm$0.02 recently obtained by \cite{prasad2023} by fitting the continuum flux densities up to 340$\,$GHz. We also identify the central 'dark lane' caused by the lack of free-free emission in a neutral circumstellar disk \citep{danchi2001} and the asymmetry in flux between the northern (brighter) and southern (fainter) lobe (see right panels in Fig.$\,$2 of \citealt{tafoya2004}). The position angle of the disk is $\simeq$100$^{\circ}$, as reported in previous observations.
    
    \subsection{Resolved H26$\alpha$ line emission}\label{sec:line}

        In the right panels of Figure \ref{fig:continuum} we show the spectral profile of the H26$\alpha$ line emission observed with ALMA in black contours. The spatial aperture over which the spectra is integrated is 0.18$\times$0.17$\,$arcsec, which corresponds to the integration limits of the model (see Appendix \ref{sec:appmodel}). The double peak profile with maxima at $-$15 and \mbox{33$\,$km~s$^{-1}$} arises from the ionized Keplerian disk. The emission is strongly maser-amplified, especially toward the edges of the disk corresponding to the emission peaks seen in the H26$\alpha$ profile, where the amplification coherence length is the largest  \citep{planesas1992} causing the saturation of the maser \citep[][see Appendix \ref{sec:appsaturation}]{baezrubio2013,tran2021}. The amplification is very sensitive to the electron density and temperature. Therefore, the asymmetry of the peaks may be caused by clumps of different size and density of ionized materials in the disk \citep{baezrubio2013}. At larger velocities, there are high-velocity broad emission components between $-$30 and \mbox{$-$70$\,$km~s$^{-1}$} and between 50 and \mbox{80$\,$km~s$^{-1}$} that arise from the ionized wind and jet. There is additional weak line emission at \mbox{$\approx-$120$\,$km~s$^{-1}$} indicated with the arrow in Figure \ref{fig:continuum} which we will discuss in Sect. \ref{sec:newline}.
        
\begin{figure*}
            \includegraphics[width=\textwidth]{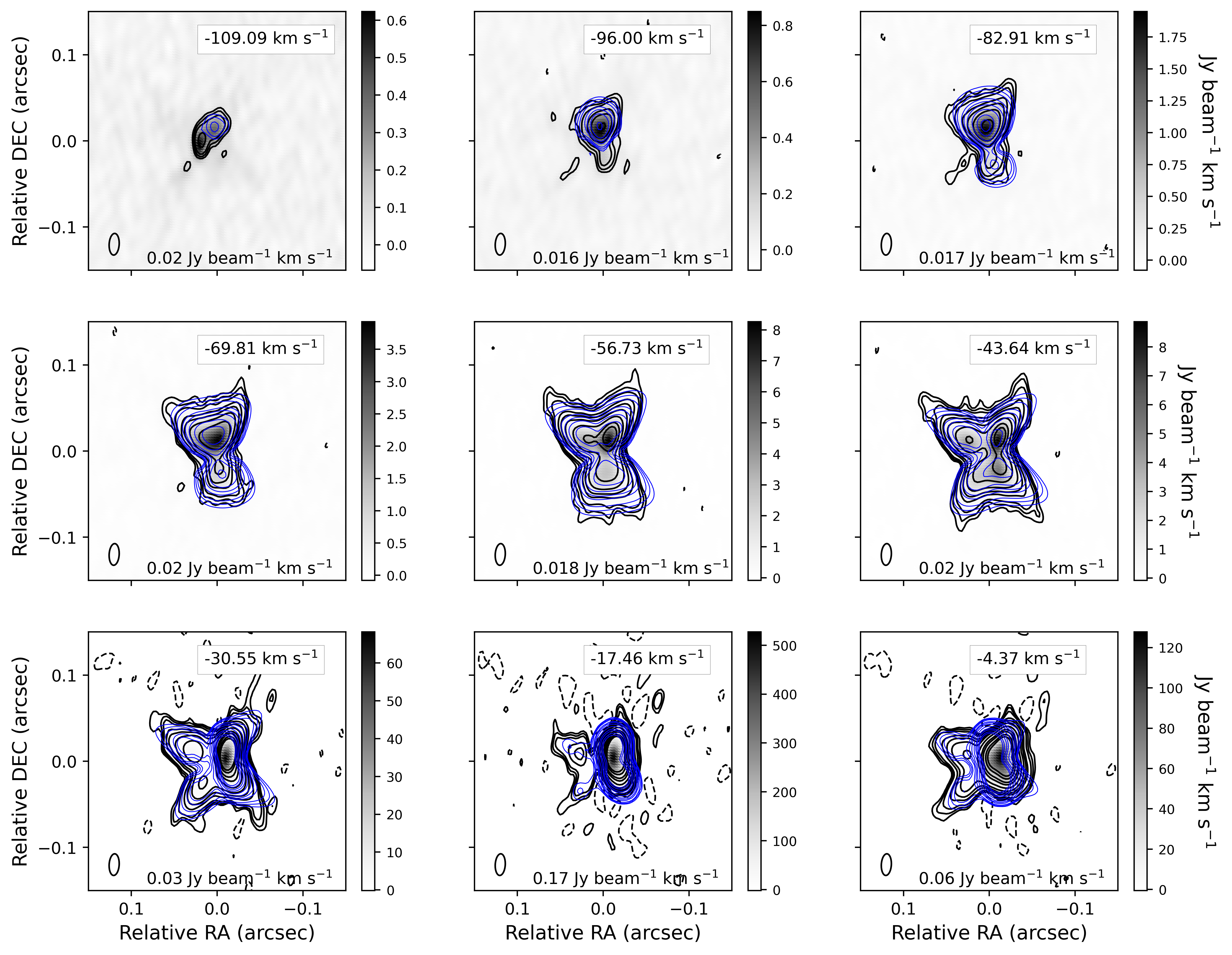}
            \centering
            \caption{Integrated intensity (moment zero) maps of the H26$\alpha$ emission in bins of 13.09$\,$km~s$^{-1}$ for the blueshifted portion of the spectrum. Gray scale and black contours represent the observed H26$\alpha$ emission; blue contours represent the emission modeled with MORELI. Contour levels are \mbox{[$-$3,5,7,10,20,30,50,100,200,300,500,700]} times the 1$\sigma$ \textit{rms} noise level of each map shown in the bottom right corner of each panel. Central velocities of each panel are indicated in their upper right corner. The synthesized beam is shown with an ellipse at the bottom left corner. The central radial velocity of the source is 8.72$\,$km~s$^{-1}$. \label{fig:panelblues}}
        \end{figure*}

\begin{figure*}
            \includegraphics[width=\textwidth]{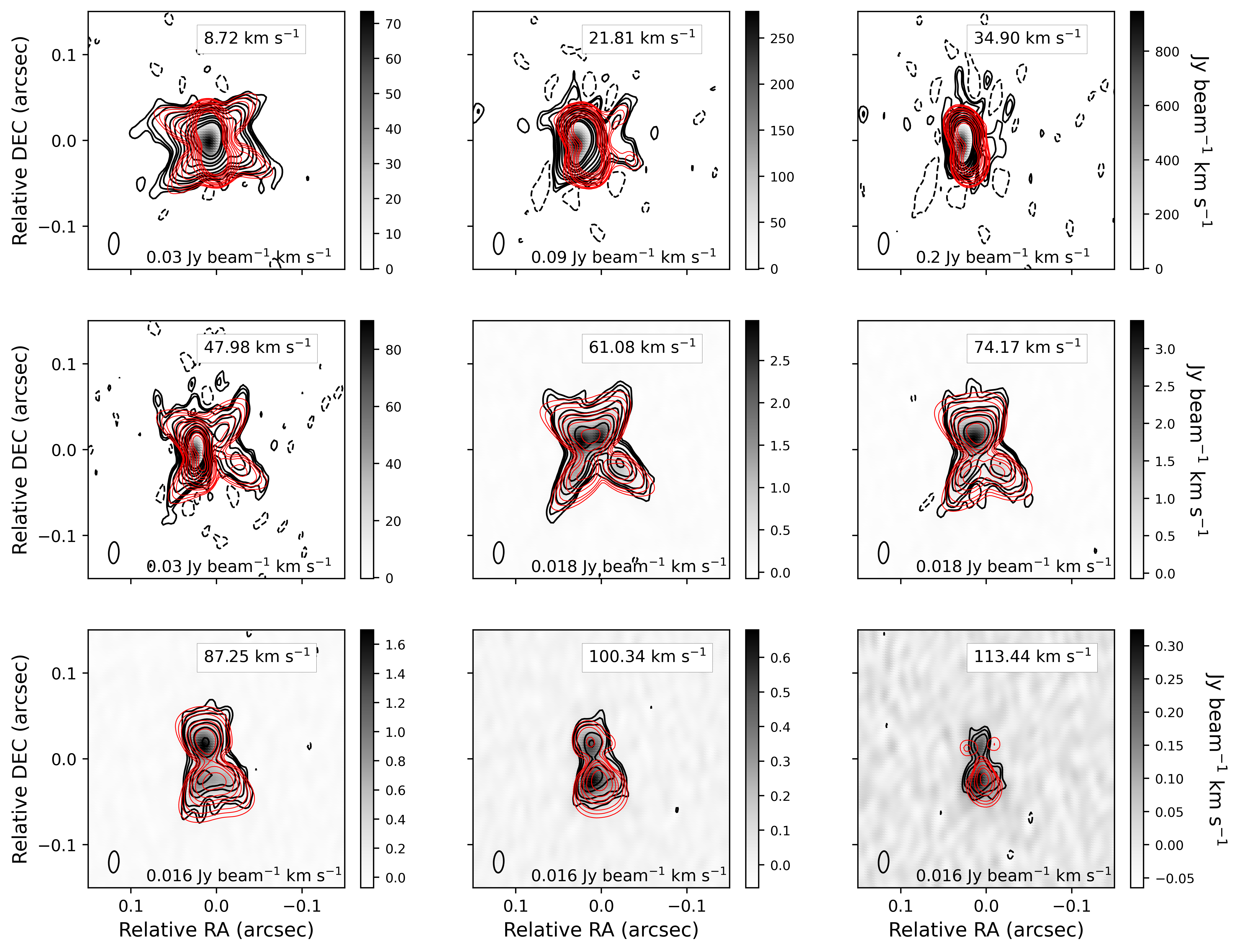}
            \centering
            \caption{Same as Fig. \ref{fig:panelblues} but for the redshifted portion of the spectrum of H26$\alpha$. The modeled images are shown with red contours. Note that the top left panel is centered at the systemic velocity of the source, \mbox{8.72$\,$km~s$^{-1}$}. \label{fig:panelreds}}
        \end{figure*}

\begin{figure*}
            \centering
            \includegraphics[scale=0.45]{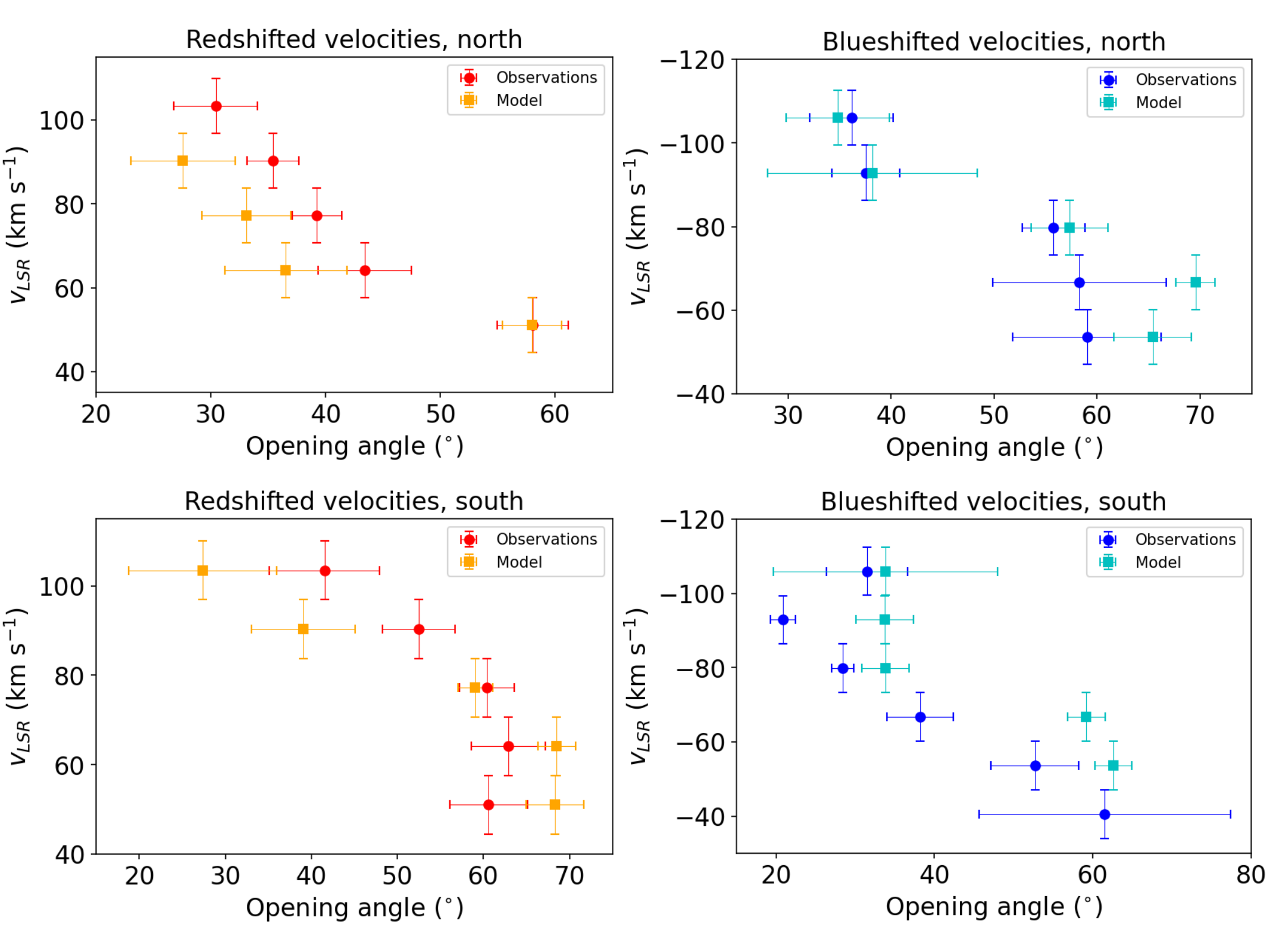}
            \caption{Opening angle of the integrated intensity maps from Figures \ref{fig:panelblues} and \ref{fig:panelreds} for modeled and observed maps, separated by velocity range and by northern and southern lobe emission. The error bars in the velocity axis represent the width of the integrated intensity maps, 13.09$\,$km~s$^{-1}$.
            \label{fig:jetopening}}
        \end{figure*}

        The ALMA images with the spatially resolved H26$\alpha$ line emission are presented in Figures \ref{fig:panelblues} and \ref{fig:panelreds}, where we show integrated intensity maps in bins of \mbox{13.09$\,$km~s$^{-1}$} for the blueshifted and redshifted emission, respectively. We calculate a systemic velocity of the source of \mbox{8.72$\,$km~s$^{-1}$} as the central velocity between the peaks of the H26$\alpha$ line. The higher velocity emission (e.g. panels with central radial velocities \mbox{$\leq-$56.73$\,$km~s$^{-1}$} and \mbox{$\geq$74.17$\,$km~s$^{-1}$}) shows a bipolar shape in the north-south direction perpendicular to the plane of the disk, which is consistent with a jet whose projection in the plane of the sky lies directly on the rotation axis. As expected for a tilted jet with the northern lobe of the jet facing the observer and the southern part moving away \citep{martinezhenares2023}, the images show stronger emission in the north at high blueshifted velocities (panels at $-$96.00, $-$82.91 and \mbox{$-$69.81$\,$km~s$^{-1}$} in Fig.$\,$\ref{fig:panelblues}) and in the south at the most redshifted ones (panels at 100.34 and 113.44$\,$km~s$^{-1}$ in Fig.$\,$\ref{fig:panelreds}). The lobes have a larger opening angle as velocity decreases, which is expected from a collimated jet whose velocity is higher near the jet axis \citep{matznermckee1999,martinezhenares2023}. 
        This effect is better seen in Figure$\,$\ref{fig:jetopening}, where the opening angle of the high velocity emission decreases for increasing velocities (red and dark blue points). The opening angle is obtained from the major and minor axes of a 2D Gaussian fit in the image plane to the integrated intensity maps reported in Figures \ref{fig:panelblues} and \ref{fig:panelreds} using CASA's \textit{imview}. 
        
        At lower radial velocities (\mbox{$\geq-$43.64$\,$km~s$^{-1}$} in Fig.$\,$\ref{fig:panelblues} and \mbox{$\leq$61.08$\,$km~s$^{-1}$} in Fig.$\,$\ref{fig:panelreds}), the emission displays the 'X' shaped morphology seen in the continuum, which is consistent with the wide-angle wind and the slower parts of the jet expanding inside the double cone geometry proposed for MWC$\,$349A \citep{martinpintado2011,baezrubio2013,baezrubio2014}. The H26$\alpha$ emission at the central velocities (\mbox{$\geq-$30.55$\,$km~s$^{-1}$} in Fig.$\,$\ref{fig:panelblues} and \mbox{$\leq$47.98$\,$km~s$^{-1}$} in Fig.$\,$\ref{fig:panelreds}) traces the ionized disk and wind. The strong, compact emission that shifts from west to east along the midplane for increasing radial velocities arises from the maser spots in the Keplerian disk. This emission corresponds to the bright double peaks of the line in Fig.$\,$\ref{fig:continuum}. The low-level emission of the H26$\alpha$ line arises from the wind, which retains the 'X' shaped morphology seen at higher velocities. The northern half is consistently brighter along the whole blueshifted (Fig.$\,$\ref{fig:panelblues}) and most of the redshifted emission of H26$\alpha$, except for the highest velocities where the southern lobe of the jet is expanding away from the observer (panels at 100.34 and 113.44$\,$km~s$^{-1}$ Fig.$\,$\ref{fig:panelreds}). This may be explained by the disk being slightly inclined from the edge-on orientation, with the northern half facing the observer (see Section \ref{sec:model}). Finally, we note a bright compact feature to the southeast in panel -109.90$\,$km~s$^{-1}$ of Figure$\,$\ref{fig:panelblues}, which corresponds to the additional emission noticed in the bottom right panel of Figure$\,$\ref{fig:continuum} (see black arrow) that we analyze in Section$\,$\ref{sec:newline}.

    \subsection{Additional line emission} \label{sec:newline}

    \begin{figure*}[]
            \includegraphics[width=\textwidth]{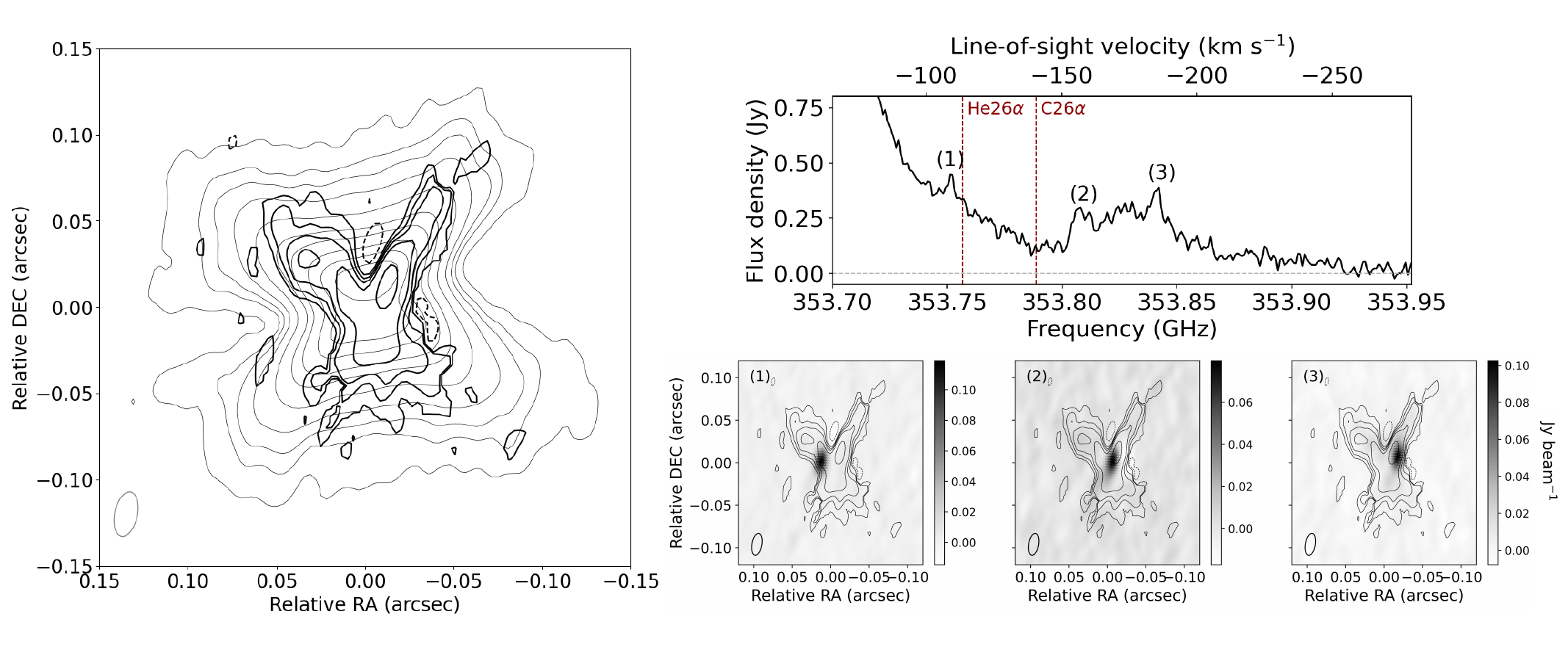}
            \centering
            \caption{\textit{Left:} Continuum emission from the line-free channels of the line spectral window in thin contours: levels are [10,20,30,50,100,200,300,500,700,900] times the 1$\sigma$ \textit{rms} of the map, 0.18$\,$mJy~beam$^{-1}$. In thick contours, the integrated intensity (moment zero) map of the line emission between 353.74 and 353.90$\,$GHz, with the contribution from H26$\alpha$ removed. Contour levels are [$-$3,3,5,7,10,20] times the 1$\sigma$ \textit{rms} of the map, 6.5$\,$mJy~beam$^{-1}$km~s$^{-1}$. \textit{Top right:} Integrated spectral profile of MWC$\,$349A showing faint emission in three peaks, (1), (2) and (3). The peaks are partially blended with the blueshifted wing of H26$\alpha$. Red dashed vertical lines represent the frequencies of He26$\alpha$ and C26$\alpha$ at the radial velocity of MWC$\,$349A, \mbox{8.72$\,$km~s$^{-1}$}. The line of sight velocity in the top axis is obtained with respect to the rest frequency of the H26$\alpha$ line. \textit{Bottom right:} In grayscale, channel maps at the frequencies of peaks (1), (2) and (3). Contours are the thick contours from the left panel. \label{fig:newline}}
    \end{figure*}

        In the top panel of Figure$\,$\ref{fig:newline} we show a zoom in of the spectral profile displaying faint emissions superimposed on the blueshifted wing of H26$\alpha$ (Fig.$\,$\ref{fig:continuum}). After removing the contribution from H26$\alpha$ pixel by pixel, we generated an integrated intensity map between 353.74 and 353.90$\,$GHz (left panel in Fig.$\,$\ref{fig:newline}) that shows the 'X' shaped morphology, suggesting that the emission arises from the ionized disk and wind of MWC$\,$349A. The emission from this new line is more compact and blueshifted than H26$\alpha$.

        The nearest lines to the frequency of the emission are the helium He26$\alpha$ and carbon C26$\alpha$ RRLs (red dashed lines in top right panel of Fig.$\,$\ref{fig:newline}). We identify peaks 1, 2 and 3 located at radial velocities of 12.58, $-$34.87 and $-$62.83$\,$km~s$^{-1}$, respectively, for the rest frequency of He26$\alpha$, while for C26$\alpha$ the radial velocities are of 39.98, $-$7.47 and $-$35.43$\,$km~s$^{-1}$ respectively. The separation between the brightest pixels for peaks 1 and 3 is 0.031$\pm$0.012$\,$arcsec, where the error is the beam size in the east-west direction. Previous works reported lines next to hydrogen RRL emission and associated them with helium RRLs at radial velocities from $-$15 to $-$40$\,$km~s$^{-1}$ \citep{thum1992,loinardrodriguez2010,martinpintado2011}.
        In our ALMA data, the peaks of H26$\alpha$ are found at radial velocities of 32.69 and $-$15.25$\,$km~s$^{-1}$ (Figure \ref{fig:continuum}), which lie closer to the radial velocities associated with the C26$\alpha$ line. The separation between the brightest pixels of the H26$\alpha$ line peaks is 0.040$\pm$0.012$\,$arcsec, similar to the new line. 
        In addition, an absorption component is found between peaks 1 and 3 that also crosses the waist of the 'X' shaped nebula from east to west for increasing frequency. Carbon RRL absorption has been previously reported at MHz frequencies from observations of the cold neutral medium \citep[e.g.][]{payne1989,oonk2017} and neutral, diffuse gas in front of HII regions \citep{salas2019}. In our case, the absorption could originate from the photo-dissociation region \citep[PDR,][]{hollenbachtielens1999}, which has a lower excitation temperature than the background HII region. Given the absorption component and the closer radial velocity to the H26$\alpha$ line, we tentatively identify this new line as C26$\alpha$, although we cannot discard some contribution from He26$\alpha$.
    
    \subsection{Comparison with the non-LTE radiative transfer model} \label{sec:model}

\begin{table*}[ht!]
        \caption{Parameters for the model of MWC$\,$349A, described in detail in Appendix \ref{sec:appmodel}. The notes in the third column indicate the method and the work where the parameters were constrained: (1) \cite{baezrubio2013}; (2) \cite{martinezhenares2023}. “Radio-continuum” refers to the simultaneous fitting to multiple radio-continuum images at different frequencies \citep{baezrubio2013}. $\theta_{jet,north}$ and $\theta_{jet,south}$ are measured relative to the rotation axis of the system, $\theta_i$. A sketch of the geometry of the model including all angles is presented in Figure$\,$\ref{fig:sketch}. \\ $^a:N_e(r=10$ au$,\theta=\theta_a)=1.6\times10^9\,$cm$^{-3}$.
        } \label{tab:allparameters}
    \begin{center}
    \begin{tabular}{c c c}
         \hline
         \hline
         Unaltered parameters &  Value & Constrained from \\  
         \hline
         
         Systemic velocity  & 8$\,$km~s$^{-1}$ & H30$\alpha$ line profile$^{(2)}$\\
         
         Central mass, $M_*$ & 23$\,$M$_{\odot}$ & H30$\alpha$ line profile$^{(2)}$ \\ 

         Keplerian ionized disk radius, $r_K$ & 46$\,$au & H30$\alpha$ line profile$^{(2)}$ \\
         
         \hline

         Density distribution, $N_e(r,\theta)^a$ & $1.6\times10^9 \left(\frac{r}{10\text{au}}\right)^{-2.14} e^{\frac{-(\theta_a-\theta)}{20}}$$\,$cm$^{-3}$ & Radio-continuum$^{(1)}$ \\

         Double-cone's semi-opening, $\theta_a$ & 57$^{\circ}$ & Radio-continuum$^{(1)}$ \\
         
         Opening angle of the ionized disk, $\theta_d$ & 13$^{\circ}$ & H30$\alpha$ line profile$^{(2)}$ \\

         Ionized disk's electron temperature, $T_d$ & 9450$\,$K & H30$\alpha$ line profile$^{(1)}$ \\

         Ionized wind's electron temperature, $T_0$ & 12000$\,$K & Radio-continuum and H30$\alpha$ line profile$^{(1)}$ \\
         
         \hline
         
         Wind terminal velocity, $v_{wind}$ & 60$\,$km~s$^{-1}$ & H30$\alpha$ centroids and line profile$^{(1)}$\\

         Keplerian rotation factor, $\alpha$  &   0.5 & H30$\alpha$ and H26$\alpha$ centroids$^{(2)}$ \\
         
         Outflow deceleration parameter, $b_v$ & $-$0.05 & H30$\alpha$ and H26$\alpha$ centroids and line profile$^{(2)}$ \\
         
         Jet peak velocity, $v_{jet}$ & 250$\,$km~s$^{-1}$ & H30$\alpha$ and H26$\alpha$ centroids and line profile$^{(2)}$ \\
         
         Jet collimation, $\psi_0$ & 0.4 & H30$\alpha$ and H26$\alpha$ centroids and line profile$^{(2)}$ \\
         
         \hline
         \hline

         Parameters updated in this work  &  \begin{tabular}{c c} Previous & Updated \end{tabular}  & Constrained from \\
         \hline

         Disk inclination angle, $\theta_i$ & \begin{tabular}{c c} 8$^{\circ}$ & $-$4$^{\circ}$ \end{tabular}  & H26$\alpha$ integrated intensity maps \\

         Northern jet tilt angle, $\theta_{jet,north}$ & \begin{tabular}{c c}$-$16$^{\circ}$ &   $-$2$^{\circ}$ \end{tabular} & H26$\alpha$ integrated intensity maps \\

         Southern jet tilt angle, $\theta_{jet,south}$ & \begin{tabular}{c c}22$^{\circ}$ &   6$^{\circ}$ \end{tabular} & H26$\alpha$ integrated intensity maps \\

        \hline
        
    \end{tabular}
    \end{center}
    \end{table*}

        We employed the 3D non-LTE radiative transfer code MORELI \citep{baezrubio2013} to model the continuum and H26$\alpha$ observations. The model parameters are presented in Table$\,$\ref{tab:allparameters}, and they are described in detail in Appendix$\,$\ref{sec:appmodel}. A sketch of the geometry of the model is shown in Figure$\,$\ref{fig:sketch}. 
        The model considers that the ionized gas is located within a double cone structure with a semi-opening angle of $57^{\circ}$, and an electron density distribution with a radial dependence $r^{-2.14}$ and an angular dependence such that the density is higher near the walls of the double cone. This electron density structure is the same as the one used by \cite{baezrubio2013}, who modeled all available radio continuum observations from radio-frequencies to the mid-IR. The observed and modeled continuum emission at 341$\,$GHz are presented in Figure$\,$\ref{fig:continuum} (left panel). The proximity of the contour levels in both the model (red contours) and the observations (black contours) indicates that the electron distribution assumed in the model closely resembles the actual distribution. The total integrated flux from the model is 2.5$\,$Jy, in excellent agreement with the observed one (2.4$\pm$0.2$\,$Jy) and with previous data (Sect.$\,$\ref{sec:continuum}). As shown by \cite{baezrubio2013} (see their Figure$\,$4), the assumed electron distribution also provides an excellent match to the morphology of the continuum emission observed with the VLA in the K and Q bands at a resolution of $\sim$0.08 and $\sim$0.03$\,$arcsec \citep{tafoya2004}, comparable to the angular resolution of our ALMA 341$\,$GHz images. We note that the self similar scaling of the 341$\,$GHz continuum image with respect to the lower frequency data is expected from the electron density distribution of the model: the $r^{-2.14}$ dependence implies that the optical depth of the continuum reaches a value of $\approx$1 at smaller radii for increasing frequencies \citep{olnon1975,panagiafelli1975}.

        In previous works, the kinematics of the system were constrained from  unresolved hydrogen RRL observations by comparing the modeled and observed line profiles and 2D Gaussian centroids: i) the disk rotates following a Keplerian law around the central star \citep{martinpintado2011}; ii) the wind is launched at 24$\,$au from the central star, rotating in the same sense as the disk and expanding at a velocity of 60$\,$km~s$^{-1}$ \citep{baezrubio2013,baezrubio2014}; and iii) the system also presents a poorly collimated, high-velocity jet expanding at a maximum velocity of 250$\,$km~s$^{-1}$ that decreases to half its maximum value at a semiopening angle of 24$^{\circ}$ \citep{martinezhenares2023}.
        
        As discussed in \cite{martinezhenares2023}, one of the parameters with the strongest effect on the 2D Gaussian centroid maps and morphology of the emission is the orientation of the jet, $\theta_{jet}$, and the \mbox{disk, $\theta_i$}. In the previous model, the disk was tilted 8$^{\circ}$ from the edge-on orientation with the southern side facing the observer. This model showed an excess of emission in the southern half at velocities more blueshifted than \mbox{$\approx-$40$\,$km~s$^{-1}$}, and no emission at the northern half for redshifted velocities higher than \mbox{$\approx$100$\,$km~s$^{-1}$}. In addition, the southern lobe was brighter for radial velocities more redshifted than \mbox{$\approx$80$\,$km~s$^{-1}$}, while in the observations this is seen from \mbox{100.34$\,$km~s$^{-1}$} (see the last two panels of Fig. \ref{fig:panelreds}).
        To overcome this mismatch with the northern and southern emission, we changed the disk orientation in steps of 1$^{\circ}$, which also modifies the wind and jet orientation (see Figure \ref{fig:thetai} in the Appendix). Reorienting the disk such that the rotation axis crossed the plane of the sky solved the issue with the asymmetrical emission seen at high velocities. By visual examination of the results, we adopted a final $\theta_i$ value of $-$4$^{\circ}$ (see rightmost panels in Figure$\,$\ref{fig:thetai}).
        
        As a second step, we updated the tilt of the jet following the same method. In the previous model the jet was tilted 16-22$^{\circ}$ with respect to the plane of the disk, with the northern cone facing the observer and the southern cone facing away. This large tilt was suggested to be caused by an unresolved circumbinary companion and/or a warped disk \citep{martinezhenares2023}. The new model requires a tilt of 2-6$^{\circ}$, much closer to the rotation axis than before (see Table$\,$\ref{tab:allparameters} and Figure$\,$\ref{fig:sketch}). Similar to the previous model, in the new simulations the northern cone also faces the observer while the southern cone moves away. As for the disk, a small change in the jet tilt angle is clearly reflected in the morphology of the line emission, hence the tilt of each cone had to be fitted individually \citep{martinezhenares2023}. 
        
        The comparison between the observed integrated intensity maps and the ones from the final model is presented in Figures \ref{fig:panelblues} and \ref{fig:panelreds}. The model reproduces very well the observed line emission across all velocities. Note that it also reproduces the higher collimation of the jet for increasing velocities, which is consistent with the behavior observed with our high-angular resolution ALMA data (see Figure$\,$\ref{fig:jetopening}).
      
        Finally, Figure$\,$\ref{fig:continuum} shows the modeled line profile obtained with MORELI (see red lines). The maser amplification of the H26$\alpha$ line is saturated \citep{tran2021}, which we tackle in the model using an approximation (see Appendix \ref{sec:appsaturation}). This approximation is known to overestimate the flux density at disk velocities between $-$15$\,$km$\,$s$^{-1}$ and 33$\,$km$\,$s$^{-1}$ (see right panels in Fig.$\,$\ref{fig:continuum}). In addition, the model predicts a symmetric line profile while the observed one is clearly asymmetric, which might be explained by the presence of clumps of ionized material \citep{baezrubio2013}. We note that other parameters such as the opening angle and electron temperature of the ionized disk also affect strongly the intensity of the line. However, these parameters were previously constrained by modeling the unsaturated H30$\alpha$ RRL in \citet{baezrubio2013} and \citet{martinezhenares2023}. At higher velocities ($\leq-$30$\,$km$\,$s$^{-1}$ and $\geq$50$\,$km$\,$s$^{-1}$), the model also predicts low-intensity line-wing emission with the blueshifted component being brighter than the redshifted one. We note that some discrepancies do exist between the observed and modeled line emission at these high velocities. This may be caused by differences between the real electron density distribution of the wind and the one assumed in the model (as e.g. inhomogeneities in the wind).
        
        The coexistence of a rotating jet and wide angle wind launched from the disk at several au from the central star is expected from magnetohydrodynamic (MHD) disk wind models \citep{koniglpudritz2000}, which have been suggested for this star previously \citep{martinpintado2011,baezrubio2014,zhang2017,martinezhenares2023}. \citet{martinezhenares2023} estimated that the jet mass loss rate and jet momentum rates derived from the model are consistent with those from other well-studied systems such as the jet from the massive protostar Cep A HW2 \citep[e.g.][]{curiel2006,jimenezserra2011}. The fact that both jet and wind are rotating in the same sense as the disk implies that they are removing angular momentum from the disk \citep{zhang2017}. It is however unclear whether MWC$\,$349A is a young star still accreting material from its disk (possibly in an advanced stage of formation given the poor collimation of the jet) or whether it is an evolved star as suggested by \cite[][and references therein]{kraus2020}.

\section{Summary}

    In this work, we have presented the first spatially resolved sub-millimeter RRL emission from the massive star MWC$\,$349A using the most extended configuration of ALMA. The unprecedented angular resolution of $\leq$0.025 arcsec ($\leq$30$\,$au) has allowed us to spatially resolve the H26$\alpha$ RRL and continuum emission arising from the ionized material in the inner parts of the system. We employed the non-LTE radiative transfer code MORELI \citep{baezrubio2013} to model the ALMA images, using the most recent version based on previous unresolved RRL observations \citep{martinezhenares2023}. This model considers an ionized disk in Keplerian rotation, a wide-angle ionized wind, and a poorly collimated, high-velocity jet, both rotating in the same sense as the disk and extracting angular momentum from the system. The resolved emission seen in the new ALMA observations can be explained by introducing minor modifications to the orientation of the disk and the jet, whose axis is closer to the rotation axis of the system than previously found. Our model of a high-velocity jet expanding inside the ionized wind is consistent with magneto-hydrodinamical launching models of the wind \citep{koniglpudritz2000}.

\break
\textit{Acknowledgements.} A.M.-H., I.J-.S and J.M.-P. acknowledge funding from grants No. PID2019-105552RB-C41 and PID2022-136814NB-I00 funded by MCIN/AEI/ 10.13039/501100011033 and by “ERDF/EU”. A.M.-H. has received support from grant MDM-2017-0737 Unidad de Excelencia "María de Maeztu" Centro de Astrobiología (CAB, CSIC-INTA) funded by MCIN/AEI/10.13039/501100011033. A.M.-H acknowledges grant CSIC iMOVE 23244 for the funding of a stay at the Harvard-Smithsonian Center for Astrophysics during which the results of this work were obtained. N.H. has been funded by grant No.PID2019-107061GB-C61 by the Spanish Ministry of Science and Innovation/State Agency of Research  MCIN/AEI/10.13039/501100011033. This paper uses the following ALMA data: ADS/JAO. ALMA No. 2019.1.01069.S. ALMA is a partnership of ESO (representing its member states), NSF (USA) and NINS (Japan), together with NRC (Canada), MOST and ASIAA (Taiwan), and KASI (Republic of Korea), in cooperation with the Republic of Chile. The Joint ALMA Observatory is operated by ESO, AUI/NRAO, and NAOJ. The National Radio Astronomy Observatory is a facility of the National Science Foundation operated under cooperative agreement by Associated Universities, Inc.


% \vspace{5mm}
% \facilities{}


\software{astropy \citep{astropy}, matplotlib \citep{matplotlib}, Common Astronomy Software Application (CASA) 6.6.3 \citep{casa}.
           }



\clearpage
\appendix

\section{Description of the non-LTE radiative transfer model}\label{sec:appmodel}

    \subsection{Structure and kinematics of the ionized gas}\label{sec:appparameters}

\begin{figure}[ht!]
        \includegraphics[width=0.8\columnwidth]{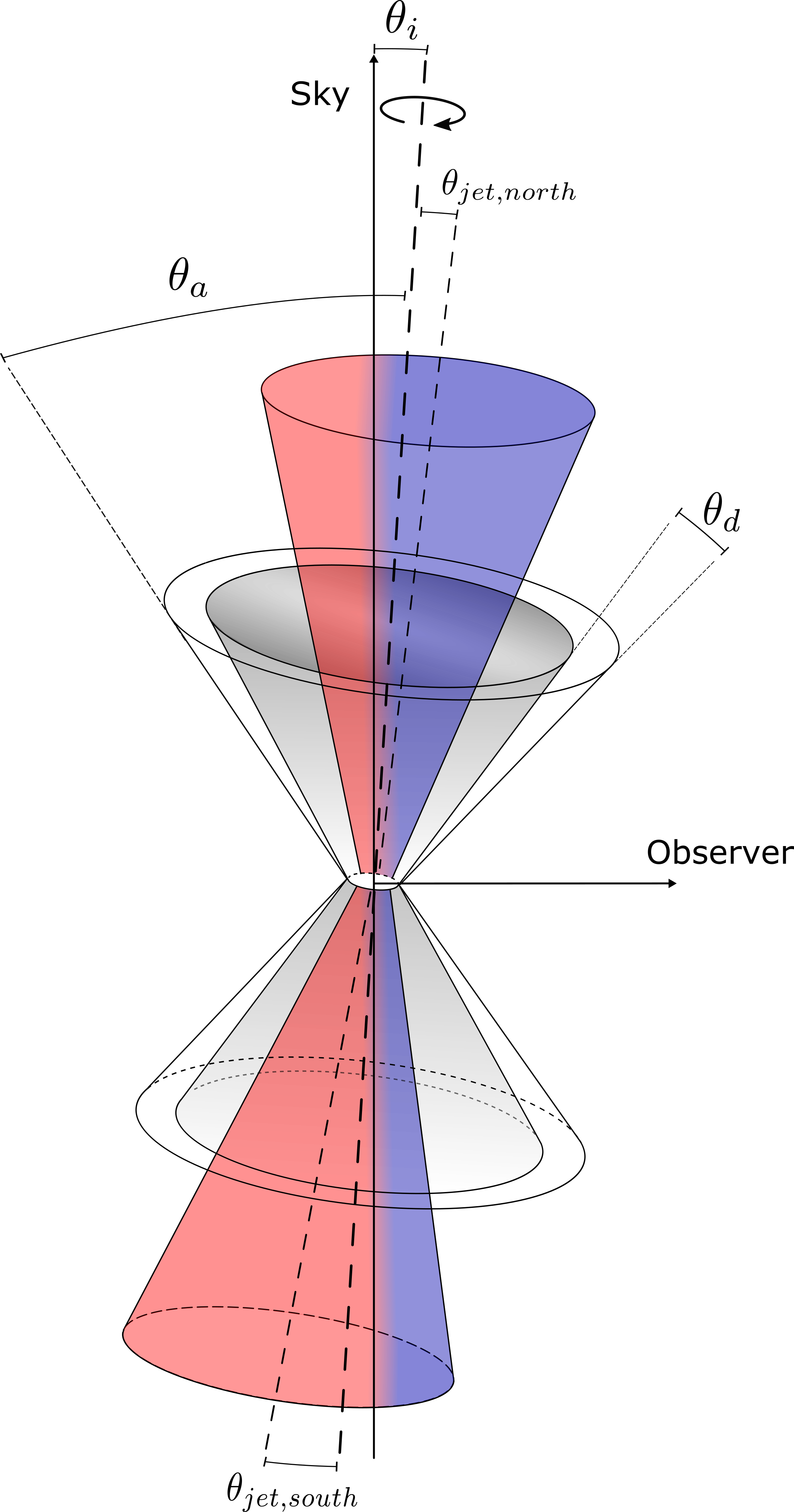}
        \centering
        \caption{Sketch - not to scale - of the geometry of the model for the ionized region of MWC$\,$349A. All angles are described in Section$\,$\ref{sec:appparameters} and the values are listed in Table$\,$\ref{tab:allparameters}. The outer cone of semiopening $\theta_a$ encloses the disk of thickness $\theta_d$ and the ionized wind (gray shaded cone) and the jet (red and blue shaded cone). The rotation axis is inclined an angle $\theta_i$ with respect to the plane of the sky. The northern jet is tilted an angle $\theta_{jet,north}$ with respect to the rotation axis, while the southern jet is tilted an angle $\theta_{jet,south}$. Blue and red colors represent blueshifed and redshifted radial velocities of the jet, respectively, as seen by the observer. Note that the jet in the model has the same size as the wind, and its shape is exaggerated in the sketch for clarity.
        \label{fig:sketch}}
    \end{figure}

    In this Appendix we briefly describe the fundamentals of the non-LTE radiative transfer code MORELI \citep{baezrubio2013}, which has been used to model the 341$\,$GHz continuum and H26$\alpha$ RRL emission from MWC$\,$349A (Sect. \ref{sec:model}). The model is described in detail in \cite{baezrubio2013}, with the most updated version that includes a jet presented in \cite{martinezhenares2023}. The parameters of the model for this work are listed in Table \ref{tab:allparameters} and a sketch of the geometry is shown in Figure$\,$\ref{fig:sketch}.
    
    MORELI considers a certain 3D geometry for the ionized region, which is discretized into a mesh of regular cubes with sizes $(dx,dy,dz)$. The $z$ axis corresponds to the direction of the line of sight; the $x$ axis is the projection of the rotation axis of the region onto the plane of the sky; the $y$ axis is orthogonal to $z$ and $x$. The integration limits are computed from the effective radius that contains the free-free emission of an isotropic, partially optically thick wind expanding radially \citep{panagiafelli1975}. The radiative transfer equation is then integrated along the $z$ axis for the whole mesh, i.e. for each line of sight. The calculation accounts for possible non-LTE effects by including the LTE departure coefficients of \cite{storeyhummer1995} computed for the local value of electron density and electron temperature of each cell. The result of this integration is the emission from the recombination line and the continuum, which are respectively arranged in a line cube and a continuum image to be compared with observations.

    In the case of MWC$\,$349A, the electron density, electron temperature and kinematics of the region are modeled considering an ionized disk, wind and jet in a double cone structure of semi-opening angle $\theta_a$ whose axis of symmetry is inclined an angle $\theta_i$ with respect to the plane of the sky. The ionized disk corresponds to the edges of the double cone of thickness $\theta_d$, which is a boundary layer between a neutral disk and the ionized wind and jet that are launched from the ionized disk (see Figure$\,$\ref{fig:sketch}). The electron density of the ionized gas in the double cone is described by a distribution $N_e(r,\theta)$ that depends on the radius as $r^{-2.14}$ and has an angular dependence such that the density is higher near the walls of the double cone (see Table$\,$\ref{tab:allparameters}). The ionized disk is characterized by an electron temperature $T_d$. It has a radius $r_K$ and rotates around the central mass $M_*$ following Kepler's law. The ionized wind is located in the inner part of the double cone with a semi-opening angle $\theta_a-\theta_d$, internal to the ionized disk, and has an electron temperature $T_0$. The wind expands with a terminal velocity $v_{wind}$ decelerated radially by the value $b_v$ and rotates around the axis of the cone following a Keplerian law scaled by a factor $\alpha$:

    \begin{equation}
        \mathbf{v_{wind}} = v_{wind}\left (\frac{r}{r_0}\right)^{b_v} \mathbf{e_r} + \alpha v_{Kepler} \mathbf{e_\varphi}
    \label{eq:wind_vector}
    \end{equation}

    \noindent where $\mathbf{e_r}$ and $\mathbf{e_\varphi}$ are the radial and azimuthal unitary vectors in spherical coordinates, respectively; $v_{Kepler}=\sqrt{GM_*/\rho}$ is the Keplerian velocity at a distance $\rho$ from the rotation axis; and $r_0$ is a characteristic length used in the model.

    The jet expands radially with a maximum velocity $v_{jet}$, and is engulfed within the wind in the inner part of the double cone.  Its orientation is described with the spherical coordinates $(\varphi_{jet},\theta_{jet})$, where the azimuth $\varphi_{jet}$ has its origin on the line of sight and increases for clockwise angles as seen from the north, and the origin of the polar angle $\theta_{jet}$ corresponds to the northern part of the rotational axis of the system. The collimation of the jet is described with a normalized distribution $f(\psi)$ adapted from \cite{matznermckee1999}:

    \begin{equation}
        f(\psi) = \frac{\psi_0^2}{\sin^2{\psi}+\psi_0^2}
    \label{eq:collimation}
    \end{equation}

    \noindent where $\psi$ is the angular distance from the jet axis to any point inside the double cone and $\psi_0$ is a flattening factor for the distribution with higher values for less collimated jets \citep{martinezhenares2023}. The azimuth of the jet powered by MWC$\,$349A corresponds to the line of sight \citep{martinezhenares2023}, so the tilt angle $\theta_{jet}$ is sufficient to describe its orientation by defining positive angles away from the observer (Figure$\,$\ref{fig:sketch}). The jet term is summed to the wind one, hence the final expression for wind and jet velocity is

    \begin{equation}
    \begin{matrix}
        \mathbf{v_{wind+jet}} = [v_{wind}+v_{jet}f(\psi)]\left (\frac{r}{r_0}\right)^{b_v} \mathbf{e_r}  \\
        + \alpha v_{Kepler} \mathbf{e_\varphi}
    \label{eq:wind_vector}
    \end{matrix}
    \end{equation}

    \subsection{Saturation of the maser}\label{sec:appsaturation}

    As mentioned in Section \ref{sec:model}, the maser emission of the H26$\alpha$ RRL is saturated \citep{baezrubio2013,tran2021}. Under unsaturated maser amplification, the intensity of the emission is amplified along the line of sight following an exponential dependence with the module of the total (line and continuum) optical depth. At a certain point in the amplification along the line of sight, the amplification reaches a limit where each pump event results in the release of a maser photon, affecting the inversion of population between the transition levels. This is the point where the maser is completely saturated \citep{strelnitski1996}, from which the emission increases linearly with the optical depth. Saturation of the maser in MWC$\,$349A takes place in the edges of the ionized disk, where the electron densities are optimum for the amplification \citep{strelnitski1996} and where the longer amplification paths towards the observer are found \citep{planesas1992}. The emission from this region corresponds to the intense peaks of the H26$\alpha$ line profile (Figure \ref{fig:continuum}) and to the integrated intensity maps at the velocities of the edges of the Keplerian ionized disk (e.g. panel at central velocity -17.46 km~s$^{-1} $ in Figure \ref{fig:panelblues} and panel at 34.90 km~s$^{-1} $ in Figure \ref{fig:panelreds}).

    MORELI accounts for the saturation of the maser considering only photons coming from the same line of sight \citep{baezrubio2013}. A more realistic treatment accounts for photons coming from every direction to evaluate the population in the atomic levels, such as in the model of \cite{tran2021}. Hence, in spite of considering the saturation of the maser, MORELI still overestimates the intensity of the line, as seen in Figure \ref{fig:continuum}. This is particularly evident in the central part of the line (between -15 and 30$\,$km$\,$s$^{-1}$), whose intensity is overestimated by a factor of $\sim2$. The parameters of the model that regulate the saturation are the solid angle of the maser beam, with a value of $4\pi/\Omega$ = 60 \citep{thum1994}, and the degree of saturation over the maser beam above which the regime turns to be linear $J_{\nu}/J_{\nu, sat}=15$. This value of $J_{\nu}/J_{\nu, sat}$ is the one used in the previous model \citep{martinezhenares2023} and has remained unchanged since it roughly reproduces the intensity of the line. A higher value results in higher intensities for the modeled line. 

\section{Influence of the disk orientation on the modeled H26$\alpha$ emission}

    \begin{figure*}[hb!]
        \includegraphics[width=\textwidth]{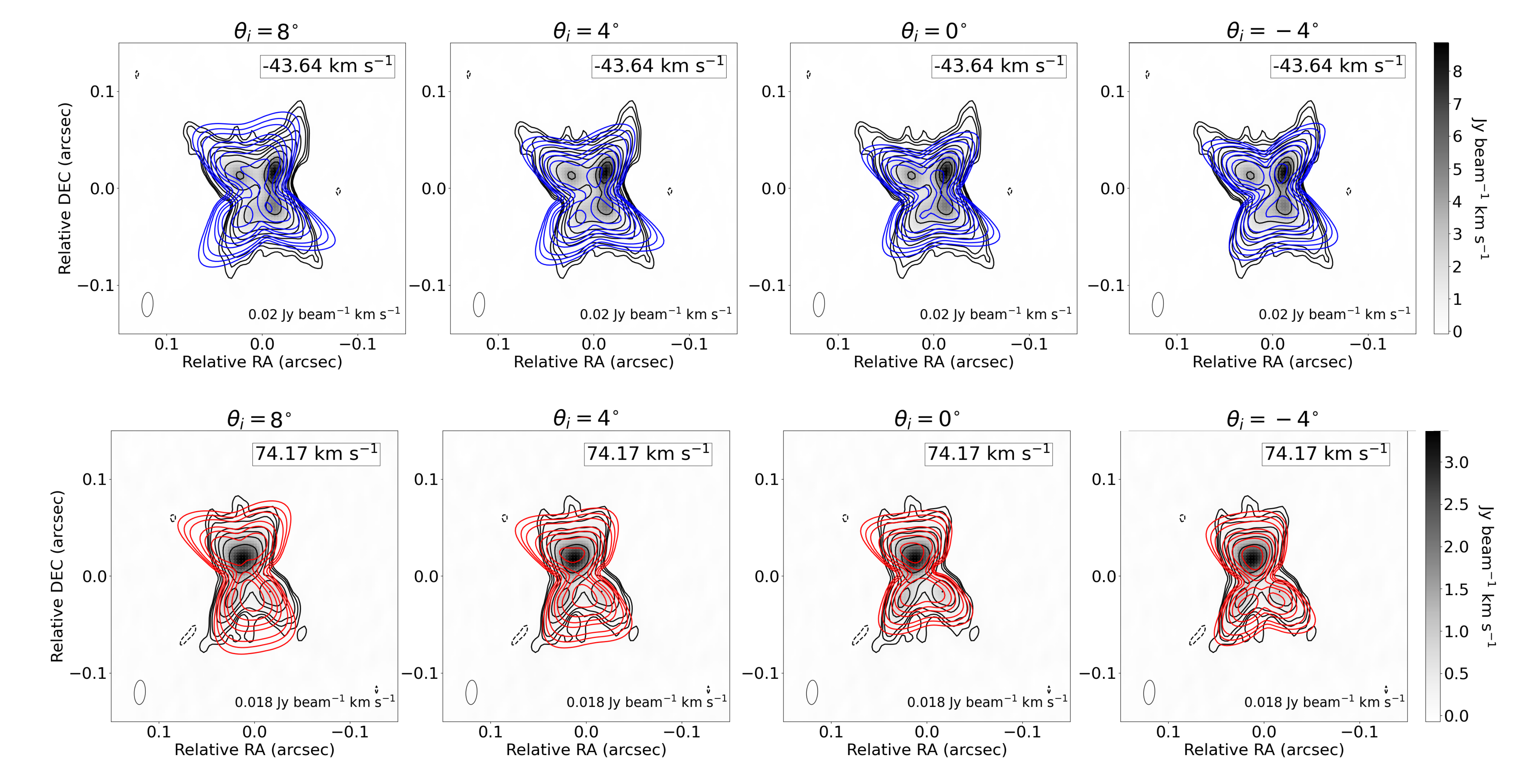}
        \centering
        \caption{Integrated intensity (moment zero) maps of the H26$\alpha$ emission from -50.19 to -37.09 km~s$^{-1}$ (top panels) and from 67.62 to 80.72 km~s$^{-1}$ (bottom panels). Grayscale, contour colors and levels and text insets are the same as in Figures \ref{fig:panelblues} and \ref{fig:panelreds}. Each column shows the model result for different inclinations of the disk $\theta_i$ with respect to the plane of the sky, where positive values of $\theta_i$ correspond to the northern cone expanding away from the observer. \label{fig:thetai}}
    \end{figure*}

    In Figure \ref{fig:thetai} we show the result of models with different values of $\theta_i$ to illustrate the effect of this parameter on the shape of the emission.
    
\clearpage
\bibliography{sample631}{}
\bibliographystyle{aasjournal}



\end{document}